\documentclass[traditabstract]{aa} 
\usepackage{graphicx}
\usepackage{natbib}
\usepackage{lscape}
\usepackage{txfonts}
\begin{document}
\def\gsim{\;\lower.6ex\hbox{$\sim$}\kern-7.75pt\raise.65ex\hbox{$>$}\;}
\def\lsim{\;\lower.6ex\hbox{$\sim$}\kern-7.75pt\raise.65ex\hbox{$<$}\;}

   \title{MAD about the Large Magellanic Cloud: \thanks {Based on observations obtained with the MCAO Demonstrator (MAD) al the VLT Melipal Nasmyth focus (ESO public data release).}}
   \subtitle{preparing for the era of Extremely Large Telescopes.}
\titlerunning{MAD about LMC}
\author{G. Fiorentino\inst{1,2}, E. Tolstoy\inst{1},
  E. Diolaiti\inst{2}, E. Valenti\inst{3}, M. Cignoni\inst{2}, \& A. D. Mackey\inst{4}}
\authorrunning{G.~Fiorentino et al.~}
\institute{
Kapteyn Astronomical Institute, University of Groningen, PO Box 800, 9700 AV Groningen, The Netherlands.\\
\email{fiorentino@astro.rug.nl}
\and
INAF- Osservatorio Astronomico di Bologna, via Ranzani 1, 40127, Bologna, Italy.\\
\and
European Southern Observatory, Karl Schwarzschild\--Stra\ss e 2, D\--85748 Garching bei M\"{u}nchen, Germany.\\
\and
Research School of Astronomy \& Astrophysics, Mount Stromlo Observatory, Cotter Road, Weston ACT 2611, Australia.\\
}

   \date{}

 
  \abstract{
We present J, H, K$_s$ photometry from the the Multi conjugate
Adaptive optics Demonstrator (MAD), a visitor instrument at the VLT,
of a resolved stellar population in a small crowded field in the bar
of the Large Magellanic Cloud near the globular cluster NGC~1928.  In
a total exposure time of 6, 36 and 20~minutes, magnitude limits were
achieved of J $\sim$ 20.5 mag, H $\sim$ 21 mag, and K$_s\sim$20.5 mag
respectively, with S/N$>10$.  This does not reach the level of the
oldest Main Sequence Turnoffs, however the resulting Colour\--Magnitude
Diagrams are the deepest and most accurate obtained so far in the
infrared for the LMC bar.  We combined our photometry with deep
optical photometry from the Hubble Space Telescope/Advanced Camera for
Surveys, which is a good match in spatial
resolution.  The comparison between synthetic and
observed CMDs shows that the stellar population of the field we
observed is consistent with the star formation history expected for
the LMC bar, and that all combinations of IJHK$_s$ filters can, with
some care, produce the same results.  We used the Red Clump magnitude
in K$_s$ to confirm the LMC distance modulus as,
$\mu_0$=18.50$\pm$0.06$_r$ $\pm$0.09$_s$ mag.  We also addressed a
number of technical aspects related to performing accurate photometry
with adaptive optics images in crowded stellar fields, which has
implications for how we should design and use the Extremely Large
Telescopes of the future for studies of this kind.
}


   \maketitle
%

\section{Introduction}

The presence of a wide variety of stellar populations of all ages
makes the Large Magellanic Cloud (LMC) a rich laboratory to trace
star-formation and evolution over a wide range of conditions and to
calibrate several primary standard-candles for distance measurements.
The LMC has been the subject of a number of ground breaking optical
imaging studies with the Hubble Space Telescope (HST) of both field
star population \citep[e.g., ][]{elson97,olsen99,holtzman99,smecker02}
and the star clusters \citep[e.g., ][]{mackey04,mackey06} of a range
of ages and metallicities.  In particular, for the first time,
accurate and detailed Colour-Magnitude Diagrams (CMDs) of star clusters
located in the crowded region of the LMC bar (e.g. NGC~1928 and
NGC~1939) were obtained \citep{mackey04}.

A recent broader, although shallower, view comes from the wide field
survey of \citet{harris09}. The global star formation history (SFH) of
the LMC is broadly consistent with that of its globular clusters.
They show that the population of the LMC has an ancient component
($\gsim$ 12 Gyr old), which is followed by a quiescent period with
very little star formation, after which there was a global episode of
star formation which started about 5~Gyr ago. This last star formation
peak may have been caused by an interaction with SMC, and this
intermediate age episode represents the bulk of the observed stellar
population. These wide field results are also broadly consistent with
deep HST CMDs (e.g., Holtzman et al. 1999, hereinafter H99; Olsen et
al. 1999).

One of the main motivations for this study is that Extremely Large
Telescopes (ELTs) are likely to be infrared (IR) optimised, using
Adaptive Optics (AO) based instrumentation. This means that sensitive
high-resolution ground-based imaging will only be possible at
wavelengths starting from optical I-band, with a peak efficiency in
the near-IR.  Both sensitivity and spatial resolution are important
for the study of resolved stellar populations, especially for compact
galaxies and also for distances beyond the Local Group. Hence it is
valuable to carry out pilot studies in this wavelength range with AO
instruments available today.

Near-IR photometry does have several advantages: it can limit the
effects of high and/or variable extinction in or towards a stellar
field and it can also provide enhanced temperature sensitivity in a
CMD, in particular when combined with optical bands. The optical-IR
colour range stretches out most of the evolutionary features in a CMD
making a unique interpretation more straight forward. However the use
of only near-IR and I filters still needs to be properly
investigated. At present stellar evolution theory is somewhat more
uncertain in the near-IR, mostly due to the atmospheric models, and to
a lack of accurate tests and calibrations over a wide age and
metallicity range.

In this paper we present a data set that allows us to {\it i)} test
the feasibility of carrying out accurate photometry of faint crowded
stellar populations obtained with an MCAO system, and {\it ii)} to
test theoretical models in the near-IR for intermediate age complex
stellar populations.  A simplistic study of the SFH of the observed
field population has been performed first by using only our near\--IR
data, and then by combining them with HST/Advance Camera for Surveys
(ACS) optical data. The consistency between optical and near\--IR
results is then discussed.

\section{MAD Observations}

\begin{figure}
\includegraphics[width=8.5cm]{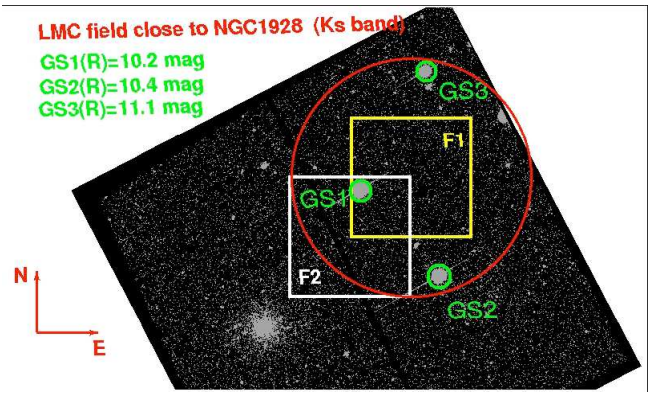}
\caption{An HST/ACS image in F555W band of the LMC field observed with
MAD.  The large (red) circle represents the entire MAD corrected field
of view. This field is only partially covered by the detector, and the two
observed fields F1 and F2 are indicated by squares.  The three guide
stars are indicated by small (green) circles (see Table 1 for details).
\label{fig-fc}}
\end{figure}

MAD, the Multi-Conjugate Adaptive Optics (MCAO)
Demonstrator\footnote{see also
http://www.eso.org/sci/facilities/develop/ao/sys/mad.html}, is a
prototype instrument built to prove the concept of MCAO using 3
natural guide stars. We were able to use it in the ``star-oriented"
wavefront sensing mode \citep[][]{marchetti06}. In this mode MAD is
equipped with three optical Shack-Hartmann wavefront sensors to
measure the atmospheric turbulence from three guide stars located,
ideally, at the vertexes of an equilateral triangle within a field of
2 arcminutes diameter.  MAD has also a ``layer-oriented" wavefront
sensing mode, built by INAF, and this mode may use up to 8 natural
guide stars, which can be fainter than the guide stars in the
star-oriented mode \citep[see ][for details]{momany08,gulli08,moretti09}. 
This mode was not offered for science demonstration observations.

The key advantage of MCAO is to increase the size and uniformity of
the corrected field of view.  MAD is able to probe the volume of
atmospheric turbulence above the telescope by performing wavefront
sensing on three natural guide stars, although successful observations
have also been achieved with only two guide stars \citep[][M.H. Wong
et al., in prep]{ferraro09}. Depending on the atmospheric seeing
conditions, the limiting magnitude of the guide stars can be as faint
as V$\sim$13 mag.  These stars drive a tomographic reconstruction of
the turbulence \citep{ragazzoni00} carried out using two deformable
mirrors conjugated at different altitudes in the atmosphere. CAMCAO is
the MAD camera equipped with a 2048$\times$2048 pixel Hawaii2 IR
detector with a pixel scale of 0.028 arcsec per pixel over a
$\sim$1~arcmin square field. CAMCAO is mounted on a movable table to
be able to scan the full 2 arcminutes field if required.  A standard
set of J, H and K$_s$ filters is available.

\begin{table}
\caption{Guide star positions and magnitudes.\label{table-guidestars}}
\begin{tabular}{lccc}
\hline
\hline
STAR & $RA_{rel}^{\prime\prime~a}$ & $DEC_{rel}^{\prime\prime~a}$ & R (mag)\\
\hline
GS1 & -19.4 & -5 & 10.2\\ 
GS2 & +15.3 & -52.5 & 10.4\\ 
GS3 & +19.6 & -50.3 & 11.1\\
\hline
\end{tabular}
\\
\tablefoottext{a}{The positions are given relative to the centre of the field 
( $\alpha$=05$^h$21$^m$11$^s$ and $\delta$ = -69$^{\circ}$27$^{\prime}$32$^{\prime\prime}$ (J2000)
)}
\end{table}

Science demonstration observations were carried out on UT3 at the VLT
during three observing runs in 2007 and 2008.  A number of studies of
crowded stellar fields (mostly Galactic) have been published using
this prototype system \citep[e.g.,
][]{bono09,ferraro09,campbell10,sana10}. They achieve a spatial
resolution that were previously only possible with space\--based
instrumentation.

MAD also provides the opportunity to understand the potential of
future AO instrumentation on large telescopes (e.g., the European
Extremely Large Telescope, E\--ELT) to perform accurate photometric
studies of distant resolved stellar populations.  A number of E-ELT
science cases assume that accurate photometry can be carried out at
very faint levels over relatively wide fields of view with MCAO
imagers.  Thus, it is important to test the potential of such systems
with currently available facilities.  The photometry of point sources
in crowded stellar fields is a useful, generally applicable, case
which provides accurate probes of photometric sensitivity and depth
over a wide field of view.  Furthermore AO currently only works
effectively at near-IR wavelengths and this is likely to remain the
case for the foreseeable future. This implies adapting current CMD
analysis techniques, which are almost exclusively carried out at
optical wavelengths.  These changes bring several challenges to be
able to interpret these images and the first step is to obtain useful
{\it training} data sets.

\begin{figure*}
\centering 
\includegraphics[width=8.5cm]{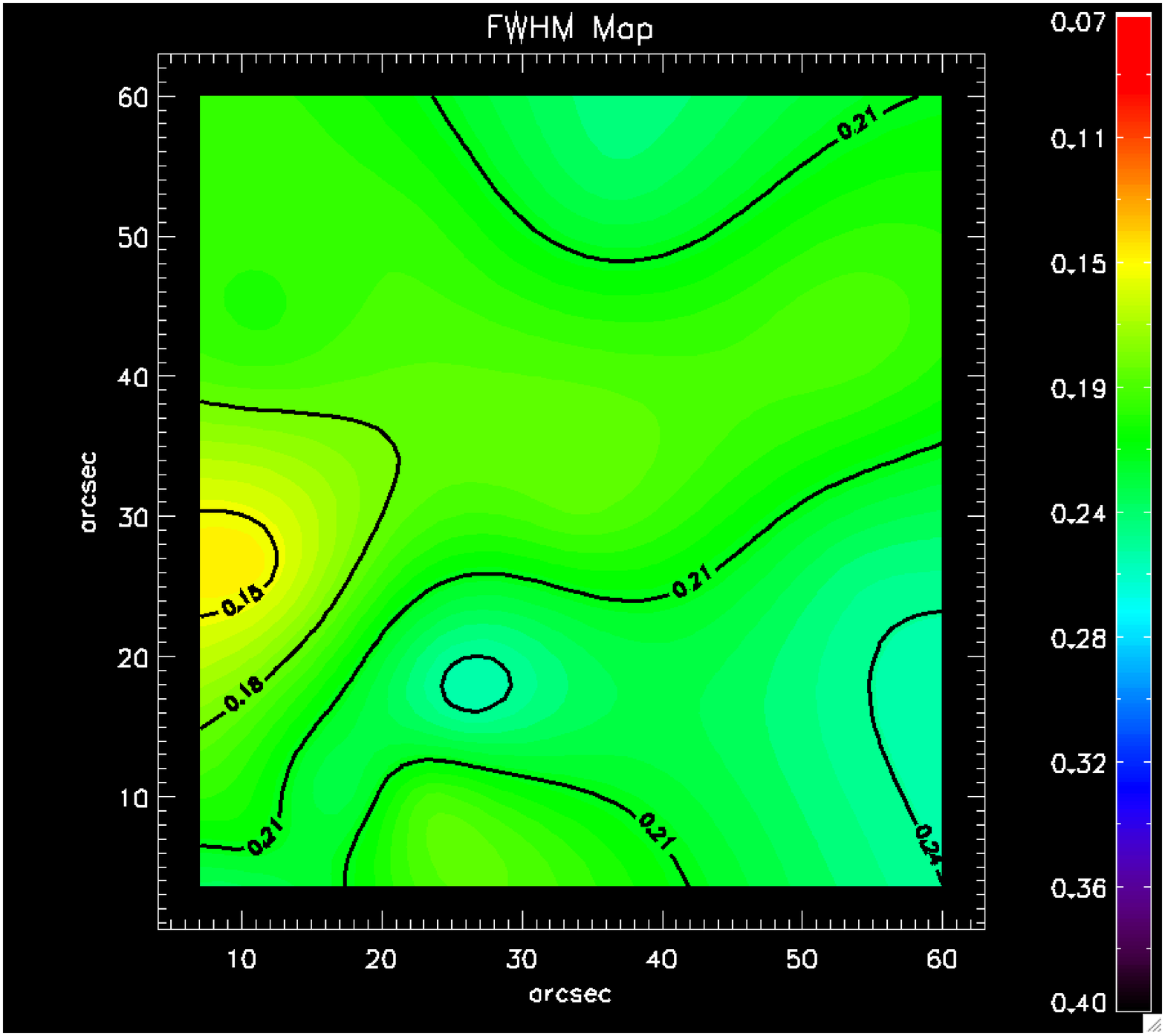}
\includegraphics[width=8.5cm]{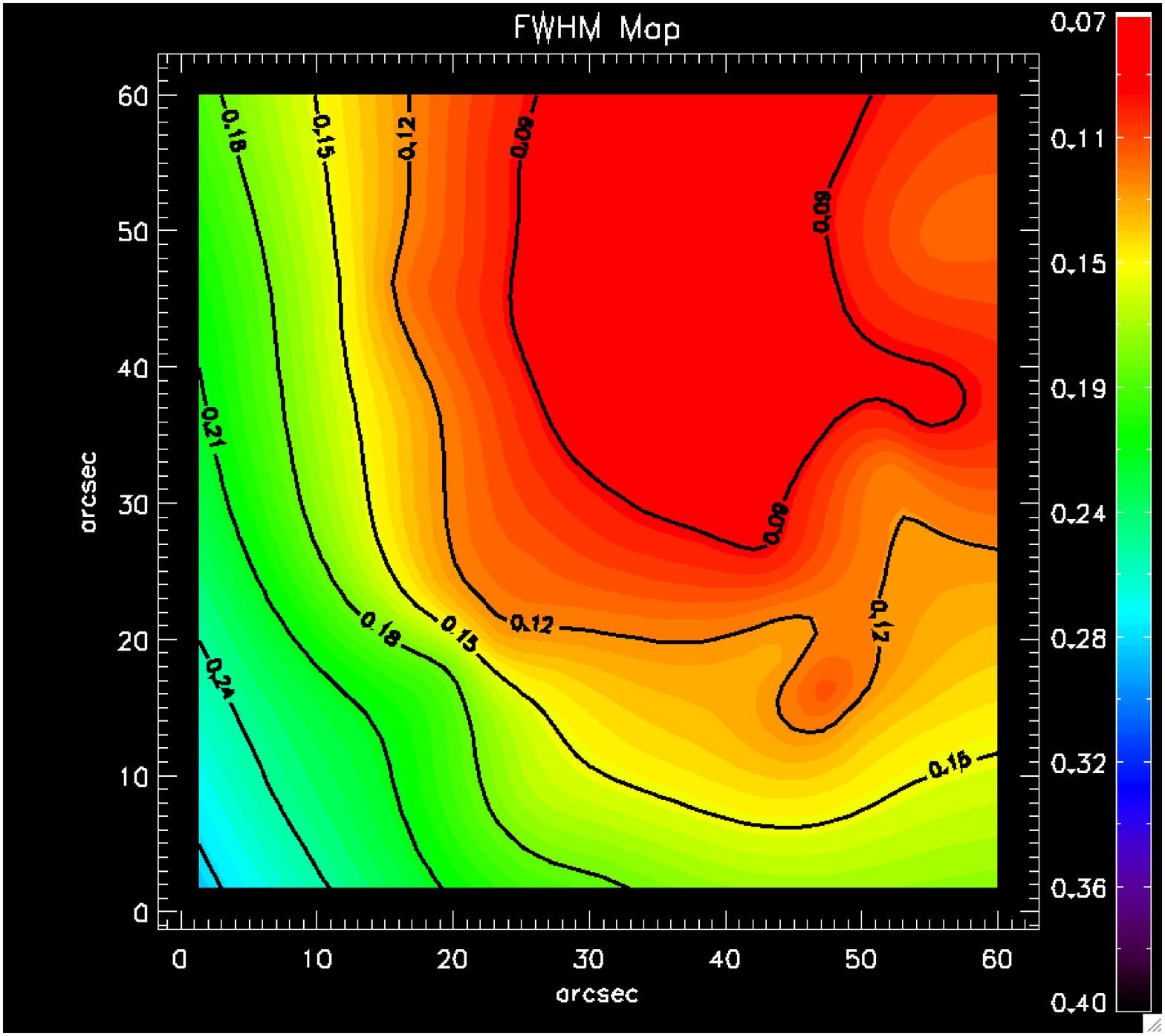}
\caption{The FWHM variation measured across the field F1 in K$_s$
filter (left) and across the field F2 in H filter (right).  The FWHM
scale ranges from 0.07$^{\prime\prime}$ to 0.5$^{\prime\prime}$, as is
shown by the colour bars on the right. The (black) contours identify
iso-density levels with a step of 0.03$^{\prime\prime}$. {\bf Left
panel:} The FWHM is quite uniform varying from 0.15$^{\prime\prime}$
to 0.24$^{\prime\prime}$. This field is well centered between the
three guide stars. {\bf Right panel:} The FWHM is very good
(0.09$^{\prime\prime}$ ) near the guide star GS1 (upper right hand
corner), but the correction rapidly declines (down to
0.24$^{\prime\prime}$ ) towards the opposite corner. 
\label{fig-fwhm}}
\end{figure*}

\subsection{MAD Data}

The MAD requirement of three bright natural guide stars within a
circle of two arcmin of diameter combined with the need to image a
region for which HST optical photometry was already available, limited
the possible sky coverage. A suitable asterism was found in a region
centered at $\alpha$=05$^h$21$^m$11$^s$ and $\delta$ =
-69$^{\circ}$27$^{\prime}$32$^{\prime\prime}$, close to the LMC
globular cluster NGC~1928. The area mapped by MAD observations is
completely covered by ACS images of NGC~1928 \citep[][]{mackey04}. For
wavefront sensing we used guide stars whose magnitudes and positions
are listed in Table~\ref{table-guidestars} and shown in
Fig.~\ref{fig-fc}.  Using this asterism we observed two $\sim1^\prime$
$\times$1$^\prime$ fields (hereafter F1 and F2).  As shown in
Fig.~\ref{fig-fc}, F1 is located at the center of the asterism and F2
is offset in a first attempt to get as close as possible to the
globular cluster NGC~1928. F2 was observed, during the first run, to
investigate the uniformity from the center to the edges of the 2x2
arcmin field of view in the three J, H and K$_s$ filters. During the
second run we observed F1 and F2 (again) only in K$_s$ filter.  We did
not manage to also observe F1 in J or H filters, and given that MAD
was a test facility that was only on the telescope for a limited time
it is not possible to rectify this. We thus present the best analysis
we can with the data available.

During the first runs (Nov. 2007 and Jan. 2008), the LMC images were
taken in J, H and K$_s$ filters following a sequence of $OSOOSOOSO$,
where S is sky and O (science) object. The final images are the
combination of 42 J, 42 H and 60 K$_s$ exposures, each 10~sec long,
resulting in the total exposure times shown in Table~\ref{table-log}.
The sky images have been taken in open loop (when the AO correction is
switched off) offsetting the telescope several arcmins away and then
applying a jitter pattern of $\sim$5$^{\prime\prime}$. During the
subsequent runs (Aug. 2008) we changed our strategy to minimize the
observing time used for the sky acquisition, by obtaining sky directly
from a series of jittered science images. A series of 60 images of
10$\times$6 sec were taken in K$_s$ band, for a total exposure time
equal to that of the previous run.

All images were processed using standard IR techniques, namely flat
field normalization, dark and sky subtraction. A summary of the
relevant technical information for all the observations is shown in
Table~\ref{table-log}. It was found that, due to rather unstable AO
conditions, only a fraction of our images were ever of a quality
suitable for photometric analysis (see section~\ref{sec-phot} for
details), e.g. in K$_s$ band this corresponds to an effective total
exposure time of about 20~min out of 60~min of open shutter time.
This means that an efficiency factor should always be taken into
account when planning deep, sensitive images with high spatial
resolution in AO\--mode.
\par

\begin{table*}
\caption{Observation log: LMC field (close to NGC
  1928).\label{table-log}}
\centering
\begin{tabular}{lccccccc}
\hline
\hline
Field ID & $RA_{rel}^{\prime\prime~a}$ & $DEC_{rel}^{\prime\prime~a}$& filter & exptime (sec) & date & $\langle$ airmass $\rangle$ &$\langle$ seeing $\rangle$\\
\hline
F1 &   0 &   0  & K$_s$ & 3600       & Aug 2008           &   1.65 &   0.45 \\    
F2 & -25 & -25  & J  & 2520       & Jan 2008           &   1.41  &   0.45 \\
F2 & -25 & -25  & H  & 2520       & Jan 2008           &   1.41  &   0.55 \\
F2 & -25 & -25  & K$_s$ & 3600  & Nov 2007 &   1.41&   0.75 \\
F2 & -25 & -25  & K$_s$ & 3600  & Aug 2008  &   1.68 &   0.65 \\
\hline
\end{tabular}
\\
\tablefoottext{a}{The positions are given relative to the centre of the field 
($\alpha$=05$^h$21$^m$11$^s$ and $\delta$ = -69$^{\circ}$27$^{\prime}$32$^{\prime\prime}$ (J2000))}\\
\end{table*}

\subsection{MAD Image Quality}

One of the major issues in making deep AO images is obtaining and
maintaining suitable image quality.  Here we quantify the variation of MAD data in terms of image quality and we investigate the possible reasons.

\begin{figure*}
\centering
\includegraphics[width=8.5cm]{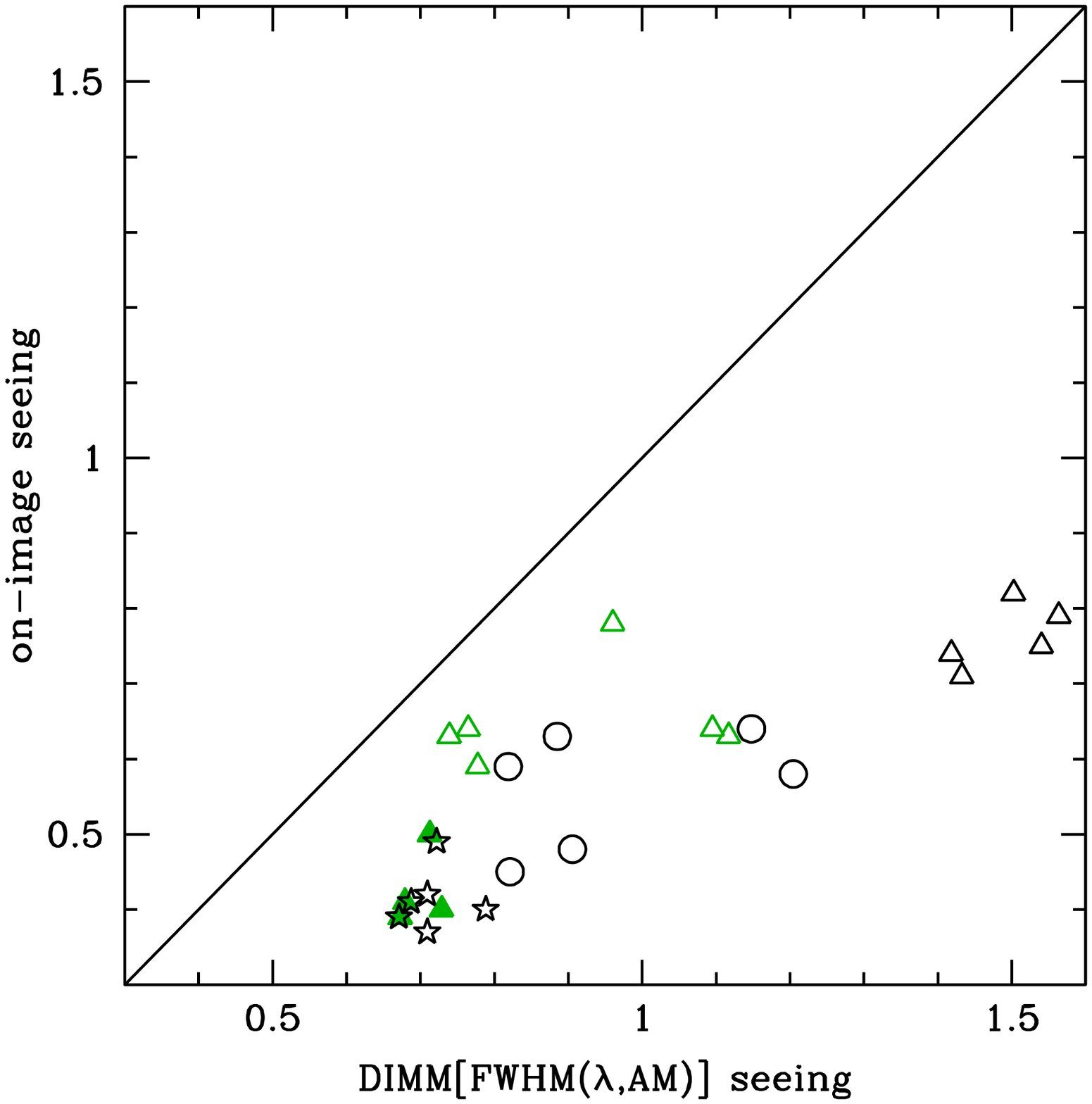}
\includegraphics[width=8.5cm]{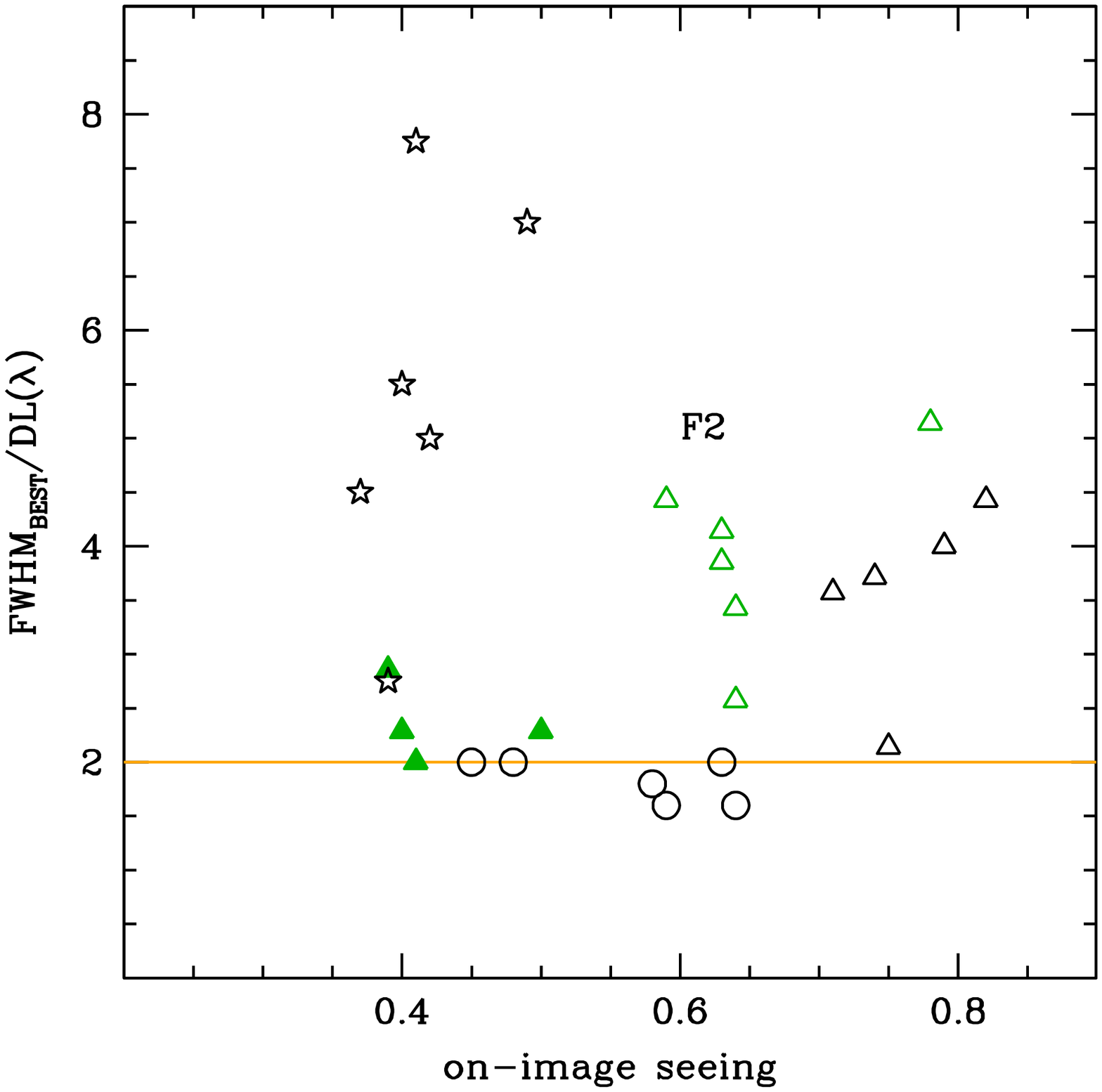}
\caption{
Triangles, dots and stars represent K$_s$, H and J measurements,
respectively. They are colour coded according to the airmass at which
the observations were taken ($\le$ 1.5 black; $>$ 1.5 green). Filled
and empty symbols represent F1 and F2 images, respectively. {\bf
Left panel:} The seeing measured by the DIMM [FWHM($\lambda$,AM)]
scaled to the actual airmass and wavelength by using the standard
relation given in the text is plotted against the FWHM measured on the
sky images taken in open loop [$\langle FWHM_{sky} \rangle$, or
on-image seeing]. {\bf Right panel:} The ratio between the best FWHM
of the Point Spread Function (PSF) 
measured in each image and the diffraction limit as a
function of on-image seeing for the same data shown in the left panel. The
solid (orange) line is 2 times the diffraction limit.
}
\label{fig-dimm}
\end{figure*}

\subsubsection{Field Position relative to Guide Stars}

We first estimated the stability of both the full width half maximum
(FWHM) and the Strehl ratio (SR) of the Point Spread Function (PSF)
across the observed fields, using an
IDL program provided by E. Marchetti.  Comparing the quality of the
correction with the diffraction limit\footnote{DL($\lambda$)
(arcsec) $\sim$ $\frac{1.22 x \lambda (cm)}{D(cm)}x$ 206265 (arcsec).
In the case of the VLT the DL is 0.04$^{\prime\prime}$ in J;
0.05$^{\prime\prime}$ in H; 0.07$^{\prime\prime}$ in K$_s$.} (DL), we
see that we can use the H and K$_s$ images of fields F1 and F2
interchangeably to study the AO performance across the MAD field of
view.  This is a reasonable approximation since the different SRs
expected for H and K$_s$ are quite well balanced by different
atmospheric conditions of our observations.

Fig.~\ref{fig-fwhm} shows how the FWHM of the PSF varies for the best image in
F1 (K$_s$, left
panel) and F2 (H, right panel).  In F1 there is a fairly uniform
FWHM distribution across the whole field in comparison to F2. The
uniformity of F1 meant that standard photometry techniques were
straight forward to apply.  In F2 however, the sharp changes at the
edges of the field were more of a challenge

Despite the strong variation over the field, the best FWHM is achieved
in F2 (0.08$^{\prime\prime}$) and it is close to the H filter
diffraction limit (0.05$^{\prime\prime}$ ). In contrast, the uniform
PSF in field F1 only reaches a FWHM of 0.14$^{\prime\prime}$, which is
twice the K$_s$ diffraction limit (0.07$^{\prime\prime}$ ). The mean
FWHM is 0.12$^{\prime\prime}$ for H and 0.20$^{\prime\prime}$ for
K$_s$. The Strehl Ratio (SR) shows a similar behaviour to the
FWHM. The SR in field F1 is quite uniform with values ranging from 5
to 15\%, whereas this distribution varies rapidly in field F2 from 5
to 25\%. The maximum SRs obtained in both fields reach, or even
exceed, the performances expected for MAD in ``star-oriented''
mode. The maximum SR was predicted to be between 11\% to 24\% for
seeing ranging from 0.7$^{\prime\prime}$ to 1.0$^{\prime\prime}$.

In conclusion, the uniformity of the PSF correction in the observed
fields changes according to the different orientation with respect
to the guide stars (see Fig.~\ref{fig-fc}).  A higher level of PSF
uniformity, over the field of view is achieved within the central area
enclosed by the asterism (in F1). However, the variations in peak
and mean FWHM between observations are mostly due to different
atmospheric conditions (see Table~2).

\subsubsection{Airmass and Seeing}\label{sec-seeing}

Many parameters, together with the target airmass and seeing, play a
crucial role in the final AO performance. The most important are the
absolute and relative magnitudes of the reference stars, the geometry
of the asterism, the spatial extent of a turbulence cell (Fried's
coherence length), the wavelength at which the correction is
performed, and the vertical distribution of the atmospheric
turbulence. As we do not have detailed information about the
atmospheric parameters (e.g., the coherence length or the turbulence
distribution) during the observations, we are only able to monitor the
performance in terms of seeing and airmass.  Hence, for a given
asterism, we compare the FWHM measured against the telescope
diffraction limit and the natural seeing.

It is well known that the seeing monitor on Paranal, the DIMM, gives
systematically higher values, which are is recorded in the image
headers, than the true seeing measured on scientific images
\citep{sarazin08}.  Since we want to accurately quantify how the
presence of the AO system affects the final images, we need reliable
estimates of the natural seeing at the time of our observations. To do
this we used the FWHM of stars measured on open loop sky images, which
are uncorrected by the AO system and thus monitor the natural seeing.
We have checked for this effect by comparing our open loop images with
the DIMM.  The DIMM sensor uses optical filter and it monitors the
atmosphere at zenith, hence we have used the standard formula
FWHM($\lambda$,AM)=DIMM(0.5/$\lambda$)$^{1/5} \times$ AM$^{3/5}$
(Sarazin 2003\footnote{VLT-SPE-ESO-17410-1174}) to derive the DIMM
value at near\--IR wavelengths and the target airmass. Fig.~\ref{fig-dimm} (left panel) clearly shows that the DIMM always
over-estimates the real seeing and the disagreement increases with
the seeing.

Fig.~\ref{fig-dimm} (right panel) shows the ratio of the best FWHM
(FWHM$_{BEST}$) in each image and the diffraction limit
(DL($\lambda$)) as a function of the on-image natural seeing.  Values
around a factor 2 of the diffraction limit (solid line in the right
panel of Fig.~\ref{fig-dimm}) are consistently achieved in K$_s$ and H
bands. Whereas in J (stars) the values are consistently much higher
than 2 times the diffraction limit. As expected, the longer the
observing wavelength, the better the AO performance. In our case the
best correction was achieved for field F2 in H band (circles). This is
probably due to the relatively low airmass of the H observations,
compared to the K$_s$.  For field F2 (and so, for a fixed asterism,
empty symbols) and constant seeing, the main difference between H and
K$_s$ images is the airmass. This airmass effect can be balanced by a
better relation between the field and the asterism. This can be seen,
that F2 H images (empty circles) show a similar correction to F1 K$_s$
images (filled triangles), because the lower airmass of the H images
compensates for the poorer AO correction.

Summarising, most of our data were obtained with an almost ideal
triangular asterism with very bright guide stars (see
Table~\ref{table-guidestars}). However it seems clear that to push the
AO capabilities to faint limits and obtain optimum results, stringent
observing constraints are needed.  It is important to observe targets
with an airmass $\lsim$1.5 and/or with a seeing $\lsim$1~arcsec.
However it should also be noted that we almost always see a
significant improvement with respect to natural seeing, of about a
factor two, even for the unfavorable airmass at which the LMC was
sometimes observed.

\subsection{MAD Photometry}\label{sec-phot}

Photometry was performed using the standard data reduction package
DAOPHOTIV/ALLSTAR and ALLFRAME provided by
\citet{stetson87,stetson94}.  DAOPHOT/ALLSTAR models the PSF by the
sum of a symmetric analytic bivariate function (typically a
Lorentzian) and an empirical look-up table representing corrections to
this analytic function from the observed brightness values within the
average profile of several bright stars in the image. This hybrid PSF
seems to offer adequate flexibility in modeling the complex PSFs that
occur in real telescopes. Furthermore, the empirical look-up table
makes it possible to account for the PSF variations (linear or
quadratic) across the chip. This gives DAOPHOT/ALLSTAR the ability to
carry out reliable photometry in many different circumstances, namely
for standard ground based telescopes, the under sampled PSF of HST and
also the complex PSFs of AO instruments.

We selected at least 100 isolated stars to estimate an analytical PSF
for each frame. These stars were chosen to cover the whole field to
account for the PSF variation. We then left DAOPHOT free to choose the
analytic function that best fits the PSF shape requiring quadratic
variations of the look-up table.  ALLSTAR was used to perform
photometry on each single frame. The resulting catalogues were used to
align with DAOMATCH and DAOMASTER all the images in our reference star
list, which is the optical catalog from HST/ACS. The near-IR images
were then processed with ALLFRAME, which performs a second PSF-fitting
by using the new, better defined, positions. ALLFRAME returns
catalogues for each image that were combined using DAOMASTER to create
the final photometric list.

\begin{figure}
\includegraphics[width=8.5cm]{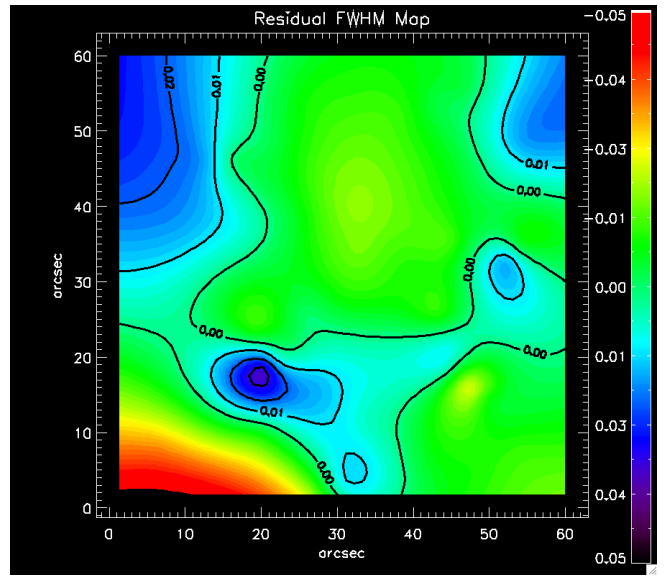}
\caption{
The residual FWHM map from subtracting the observations of F2 in H
band from the PSF model created using DAOPHOT.
\label{fig-sub}}
\end{figure}

If we compare the observed PSFs, and their variation over the
field, with the PSF models created using DAOPHOT 
we see a good agreement. This comparison is
shown in Fig.~\ref{fig-sub}, for the challenging case of F2 in H band (see
Fig.~\ref{fig-fwhm}, right panel), where the residuals of the
subtraction of the two distributions are shown.  Fig.~\ref{fig-sub}
shows how well the model follows the observed PSF distribution with
only a few exceptions. The residuals are largest in the regions
farthest from the guide stars, and where there are rapid changes in the
observed PSF.

It is worth to mentioning that the PSF modeling and fitting procedure
across the field turned out to be particularly easy thanks to the
highly uniform AO correction provided by MAD.  Accurate photometry
with data acquired with a single conjugate AO (SCAO) system requires
the use of complex highly variable PSFs, which require a combination
of several Gaussians to take account of the strong variation in PSF
shape across the image (see Origlia et al. 2008).  This difference
represents the major advantage that MCAO can provide.

The final matched photometry catalogues, containing 1218 and 1232
stars in fields F1 and F2, respectively, were obtained by
cross\--correlating the single\--band catalogs using {\it CataXcorr}
(P. Montegriffo, private communication) with {\it rms} residuals of
$\approx \pm0.2\arcsec$. There are $\sim$140 stars which have been
observed in both F1 and F2, but only in K$_s$.

The instrumental magnitudes were converted to the Two Micron All Sky
Survey (2MASS) photometric system, using 2MASS
catalogues. Unfortunately, we found few stars in common between the
two catalogues, due to the very different spatial resolution of MAD
and 2MASS. This means that the comparison has a large overall
uncertainty of $\pm 0.10$ in J, $\pm0.19$ in H and $\pm0.09$ in K$_s$ in
the zero\--point calibration.  By defining our limiting magnitude as
the faintest magnitude with a S/N $<$ 10, we reach: J $\sim$ 20.5 mag;
H $\sim$ 21 mag; K$_s$ $\sim$ 20.5 mag.

 Most of our observations were taken in variable seeing
conditions, and thus many individual images are not very sensitive due
to poor S/N. To understand how our limiting magnitude corresponds to a
total effective integration time we determined in how many images the
stars at the limiting magnitude were detected. The number of these
images times the exposure time for each image, has been considered as
our `effective integration time'. From this definition, we used only
14\% of the images in J band (corresponding to 6 mins of exposure
time), 86\% in H band (36 mins) and 33\% of the images in K$_s$ band.

\section{HST/ACS Observations}\label{sec-hst}

The HST optical observations were taken during Cycle 12 using the ACS
Wide Field Channel (WFC) centered at $\alpha$=05$^h$20$^m$57$^s$ and
$\delta$ = -69$^{\circ}$28$^{\prime}$41$^{\prime\prime}$. Two
exposures were taken through the F555W filter (330 s) and one through
the F814W filter (200 s). The HST/ACS photometry has been carried out
with DAOPHOT (in the IRAF environment) and is fully described by
\citet{mackey04}. The HST/ACS photometry is very deep (at least 4 mag
below the Main Sequence Turn Off (MSTO), i.e. V, I $\sim$ 26 mag) and
the completeness is 100\% at the level of our faintest IR
observations. This optical catalogue was combined with the JHK MAD
catalogue, creating a final optical-IR catalogue of $\sim$2600 stars.

These optical data also allowed us to simply derive the completeness
of our IR catalogue by matching the stars retrieved in the IR
photometry to the complete HST photometry. Due to the large
colour range (from V to K$_s$) of our data set, the completeness has
been derived independently for V$-$I$>0.75$ and V$-$I$<0.75$.  In the
following analysis we used only stars with V magnitude brighter than
21 where both blue (V$-$I $\leq$ 0.75) and red (V$-$I $>$ 0.75)
completeness values are higher than 50 \%, in the three near--IR
bands.

\section{Results}\label{sec-3}

\subsection{The optical/IR CMDs}

\begin{figure*} 
\centering 
\includegraphics[width=5.7cm]{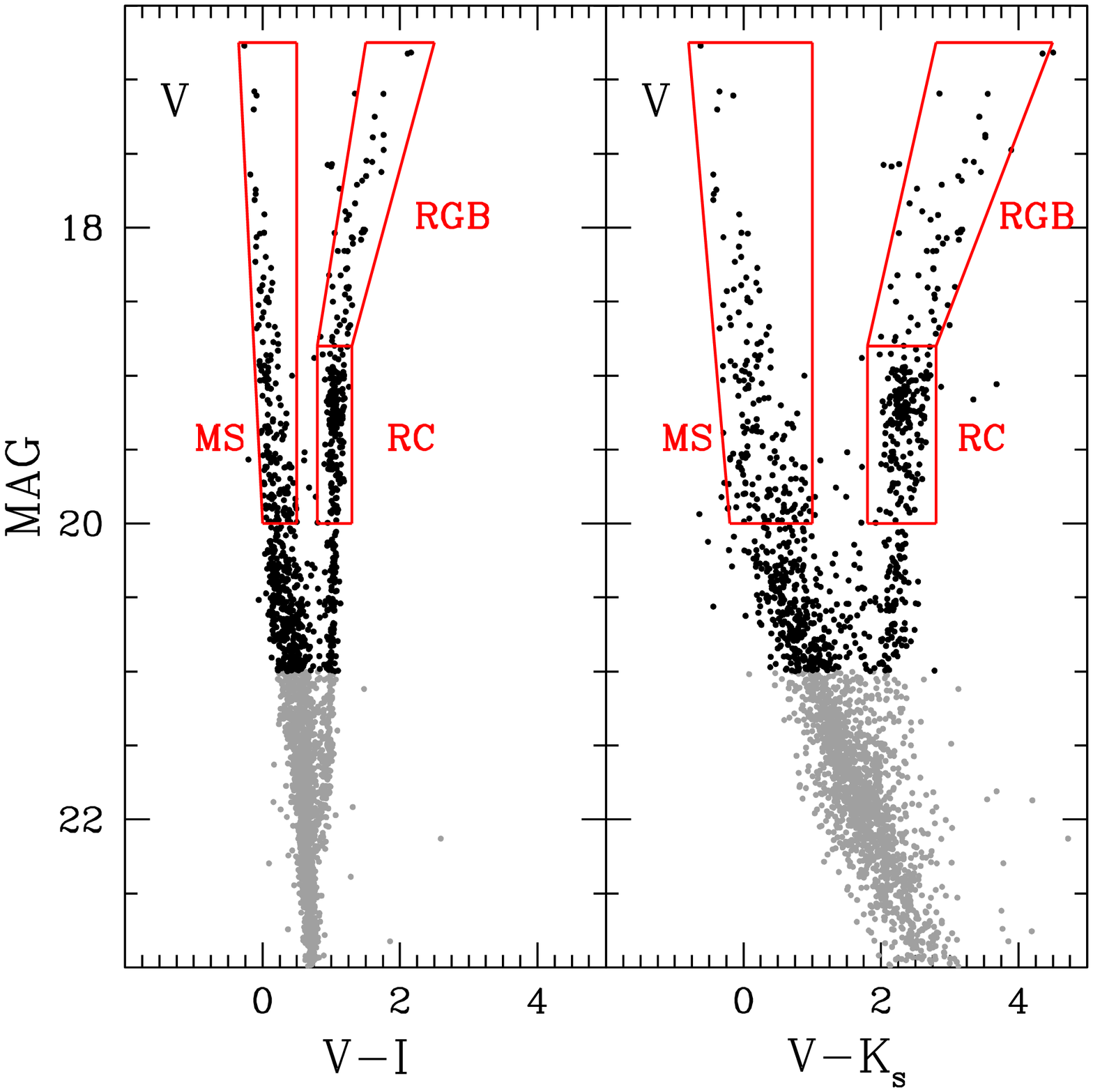}
\includegraphics[width=5.7cm]{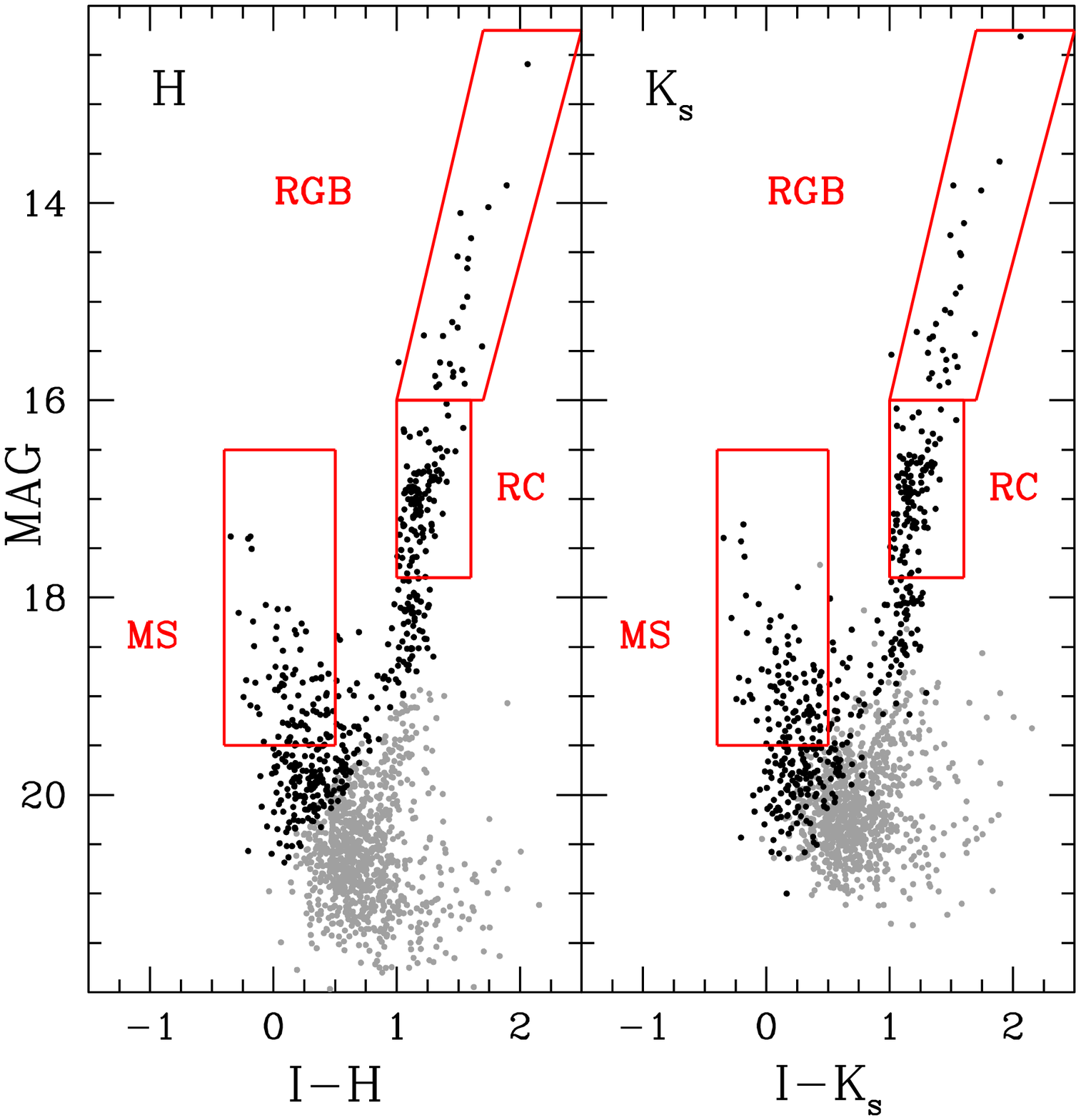}
\includegraphics[width=5.7cm]{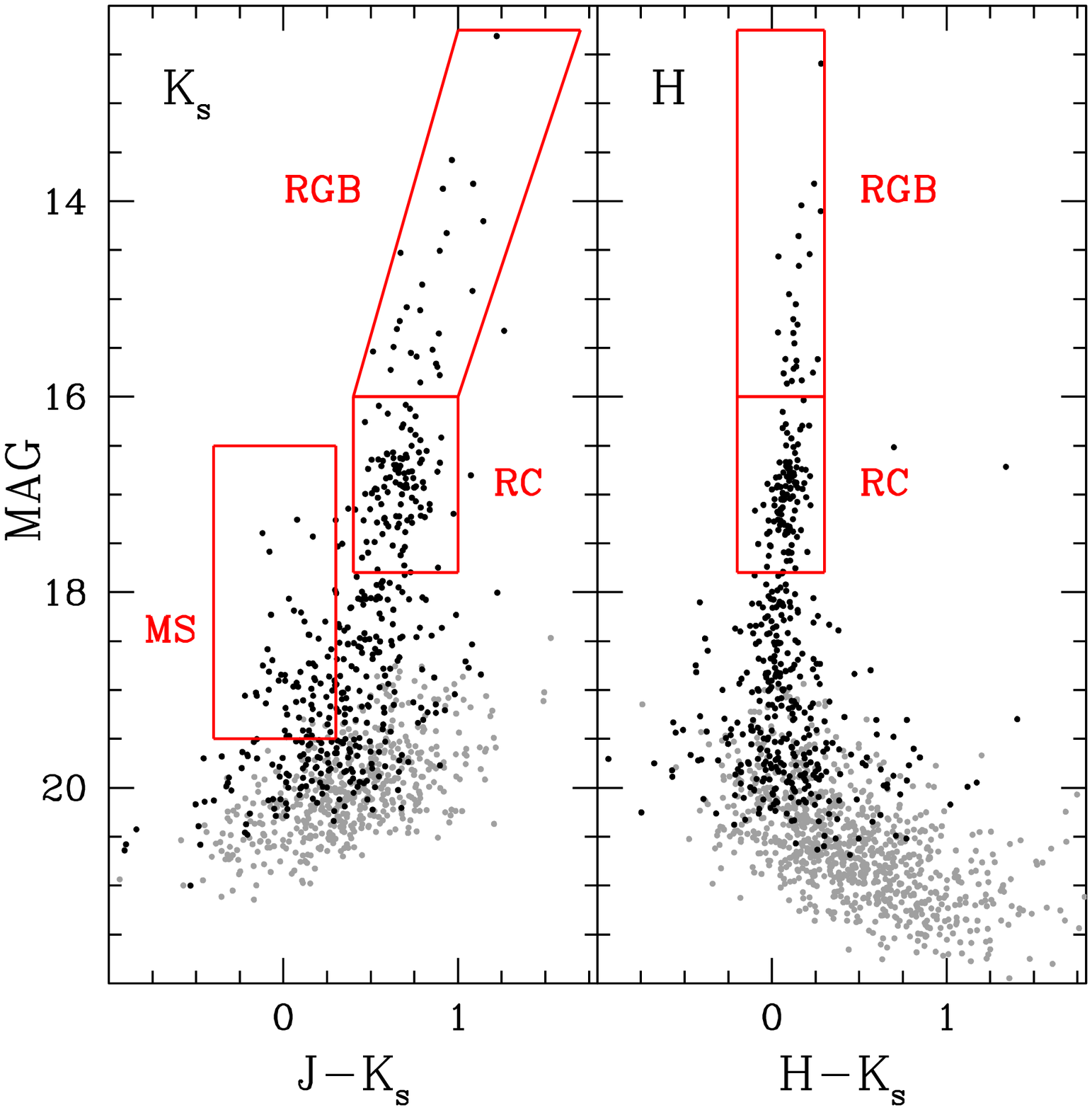}
\caption{
The Colour-Magnitude Diagrams for our LMC bar field in different
filter combinations.  Starting from optical HST/ACS V,~V$-$I (1st
panel on the left), and extending to optical-IR, by adding MAD H and
K$_s$ photometry (central panels), and finally the MAD IR photometry
alone (panels on the right). The grey points are all the sources found
by Allframe and matched between different filters, and the black are
only those with a completeness higher than 50\% (see
section\ref{sec-hst}). The boxes outline the main features of the
CMDs, the Main Sequence (MS), the Red Giant Branch (RGB) and the Red
Clump (RC).
}\label{cmd}
\end{figure*}

In Fig.~\ref{cmd} we show CMDs obtained from the optical\--IR catalog
with different filter combinations, most of which are likely to be
possible with an imager on the E-ELT \citep[e.g.,][]{deep11}.

The best defined and most extended feature in an optical\--IR CMD is
of course the Red Giant Branch (RGB), see Fig.~\ref{cmd}.  The RGB
signals the presence of stars $\gsim$1~Gyr old.  It is notoriously
difficult to interpret uniquely due to age-metallicity degeneracy
effects.  In all the CMDs in Fig.~\ref{cmd} there is also the clear
presence of the Red Clump (RC), located at V$\sim$19 mag and at H, K$_s$
$\sim$ 17 mag, which indicates the presence of intermediate and old
stellar populations, in the range 1$-$10 Gyrs old.  A blue plume shows
the presence of young Main Sequence (MS) stars ($< 100 - 500$~Myr
old). A spread of stars just above the RC, is also clearly visible in
Fig.~\ref{cmd}, and is due to the presence of young blue loop stars.
These can be useful to derive the metallicity of stellar populations
$\lsim$1~Gyr old (e.g., Dohm-Palmer et al. 1998).

As is clearly shown in Fig.~\ref{cmd}, the combination of
optical and IR filters stretches out the main CMD features, hence
allowing an easier and more precise separation of different stellar
populations in all evolutionary phases.

\subsection{The Star Formation History}

There exist robust techniques to make detailed analyses of
observed CMDs by comparing them to theoretical models
\citep[e.g.,][]{tosi91,tolstoy96a,aparicio97,dolphin02a,aparicio04,cignoni10}.
From theoretical evolutionary tracks we know that in an optical-IR CMD
the features will be stretched out due to the long colour baseline.
However, although in the optical domain the theoretical stellar
evolution models are well calibrated, in the near-IR they still have
to be fully verified for a range of stellar evolution phases. Moreover
this needs to be confirmed from an observational point of view using
the actual photometric accuracies.
 
Here we present a basic application of the well establish CMD
synthesis methods to interpret our optical/IR CMDs in terms of a
likely SFH. We start with a ``known" SFH and see
what are the effects of changing this. Given the uncertainties in the
IR stellar evolution tracks we pay special attention to comparing the
IR and IR$+$optical CMD results to the known SFH obtained with optical
photometry alone.

\begin{figure*}
\centering 
\includegraphics[width=8.5cm]{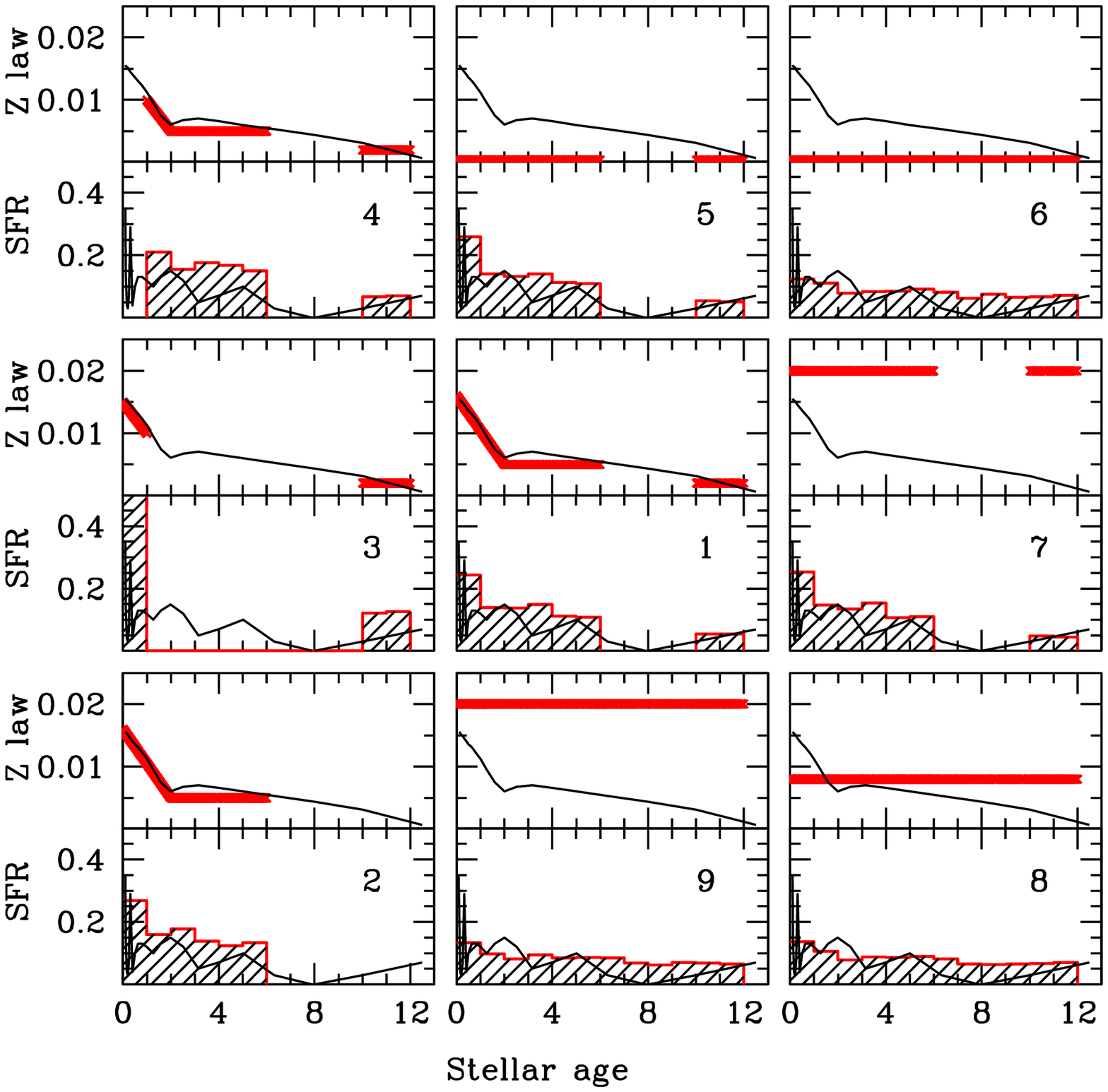}
\includegraphics[width=8.5cm]{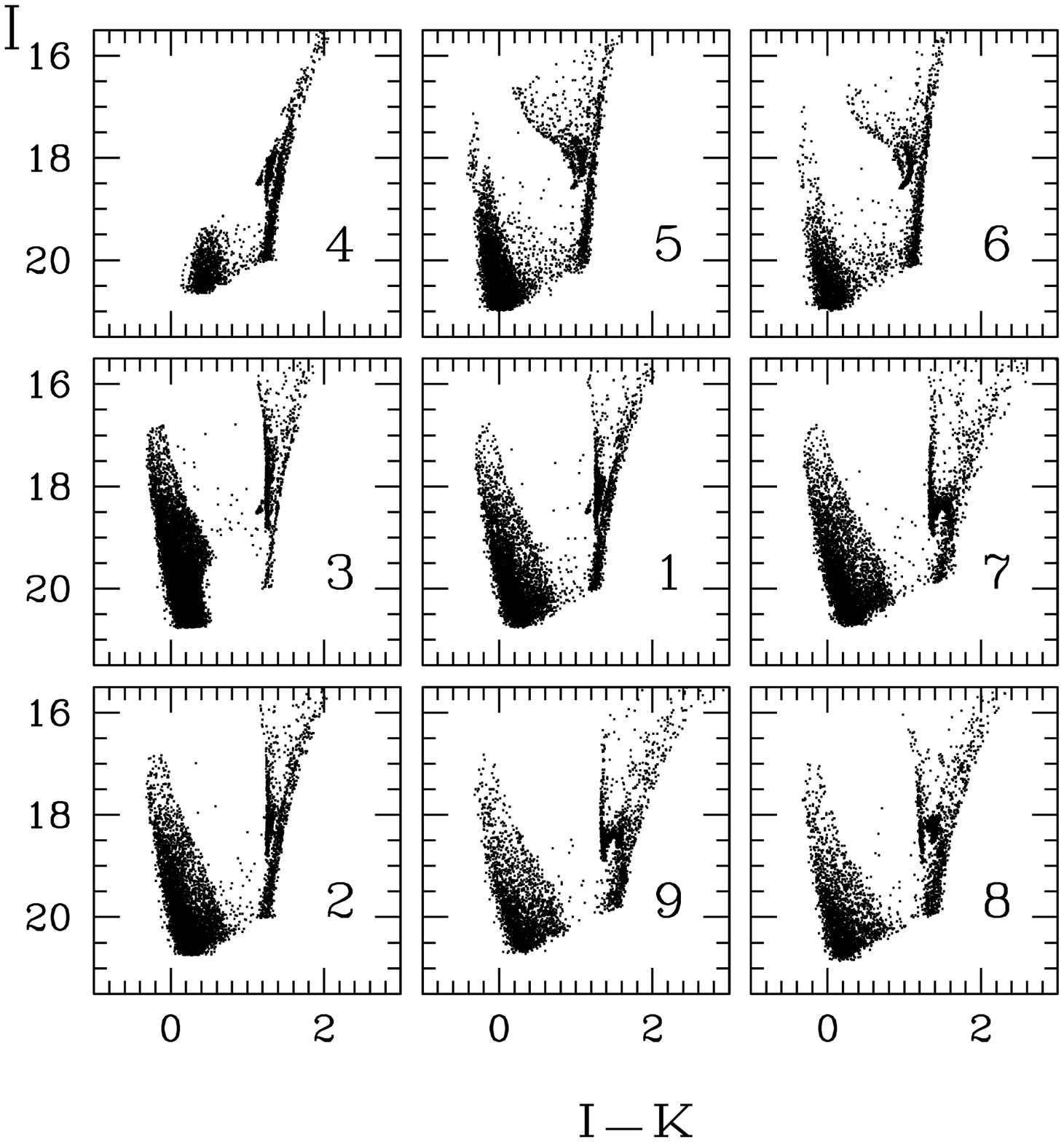}
\caption{
{\bf Left panel:} A series of synthetic SFH and metallicity functions are
shown. The thin (black) lines are always the SFH and metallicity
function of a field of the LMC bar, as derived from \citet{holtzman99}. 
The thick (red) lines and the histograms show the values
used to simulate the synthetic populations in order to find the best
model that matches our observations. {\bf Right panel:} Synthetic I,
I$-$K$_s$ CMDs for different SFHs and metallicity functions, given in left
panels.  The number of the model is specified in each
single panel as it is referred to in the text and in
Tables~\ref{table-num} and \ref{table-clump}.  
 }\label{sfh}
\end{figure*}
\vspace*{0,5cm}

Due to the limited depth and field of view of our IR photometry we
cannot make a reliable determination of the SFH directly from our IR
data. As a full SFH analysis from optical data already exists for a
nearby LMC bar field, and within the errors of this kind of analysis,
the optical CMDs are identical \citep[see H99 \&][]{mackey04}, this
allows us to directly compare our results to the careful and detailed
SFH determined by H99.

We carry out a simple comparison, counting the number of stars falling
in selected regions of the observed and theoretical CMDs for different
filter combinations (as shown in Fig.~\ref{cmd}); then we compute
the probability that the chosen theoretical model (i. e., the
synthetic CMD) matches our data.

We first recreated a synthetic population assuming the SFH and
metallicity function from H99. This is our reference model (number 1
in Fig.~\ref{sfh}, left panel). Then, we built 8 different
populations, by changing the assumptions on both metallicity and age
of the reference model. The variety of SFHs and the metallicity functions
that have been explored are shown in Fig.~\ref{sfh} (left panel) with
(red) lines and are compared to the H99 SFH (black) lines.

The models can be summarised as follows: 1) H99; 2) H99 without the
old population; 3) H99 without the intermediate age population; 4) H99
without the very young population; 5) H99 with a very metal-poor and
constant metallicity (Z=0.0001); 6) Constant star formation rate
(SFR), and constant low metallicity (Z=0.0001); 7) H99 with a
constant, high metallicity (Z=0.02); 8) Constant SFR and constant,
intermediate metallicity (Z=0.008); 9) Constant SFR and constant, high
metallicity (Z=0.02).

The whole set of synthetic populations (each with one $\sim$ 50~000
stars) have been simulated using the IAC-STAR \citep{aparicio04},
which generates synthetic CMDs for a given SFH and metallicity
function. Composite stellar populations are calculated on a
star-by-star basis, by computing the luminosity, effective
temperature, and gravity of each star by interpolation in the
metallicity and age grid of a library of stellar evolution tracks. We
used the stellar evolution libraries of \citet{girardi00}, the
bolometric corrections libraries from \citet{castelli01} and we
assumed a Salpeter Initial Mass function. We fixed the reddening to
E(B-V) = 0.075 mag and the distance modulus to $\mu_0$=18.50 mag, from
H99. In order to compare our observations to theoretical models, K$_s$
band magnitudes were transformed to the Bessell \& Brett photometric
system by adding 0.044 mag \citep[see][and reference therein for
details]{salaris03}.

\begin{table*}
\caption{The number of stars falling in
  the boxes outlined in Fig.~\ref{cmd}, for
  different combination of filters. These boxes represent: Young (MS) and intermediate-old (RC and RGB) stellar populations.\label{table-num}}
\centering
\begin{tabular}{lccccccccccccccccc}
\hline
\hline
N  & MS & RGB & RC  &MS & RGB & RC & MS & RGB & RC & MS& RGB & RC& MS& RGB & RC & RGB & RC\\
   & (V$-$I) & (V$-$I) & (V$-$I)  &(V$-$K) & (V$-$K) & (V$-$K) & (I$-$K) & (I$-$K) & (I$-$K) & (I$-$H)& (I$-$H) & (I$-$H) & (J$-$K) & (J$-$K) & (J$-$K)&(H$-$K) & (H$-$K)\\
\hline
0BS &  183  &   52   &   195 &    175  &   53 &    190  &   151  &    46 &    197 &    86  &  26  &  111 &   83 & 24  & 100  &     26  &   113\\
1   & 2055  &   291  &  1176 &   2053  &  325 &   1176  &  1609  &   270 &   1297 &  1624  & 252  & 1292 & 1740 & 266 & 1298 &    294  &  1382\\
2   & 2265  &   319  &  1195 &   2260  &  367 &   1195  &  1803  &   306 &   1337 &  1821  & 289  & 1335 & 1970 & 300 & 1337 &    317  &  1433\\
3   & 5656  &   387  &  1093 &   5656  &  464 &   1092  &  4370  &   319 &   1330 &  4353  & 269  & 1364 & 4375 & 310 & 1333 &    362  &  1546\\
4   & 54    &   209  &  1245 &     52  &  208 &   1245  &   130  &   196 &   1324 &   158  & 183  & 1321 &  310 & 196 & 1324 &    196  &  1324\\
5   & 865   &   168  &   565 &    995  &  250 &    562  &   477  &   192 &    744 &   506  & 159  &  654 &  529 & 169 & 906  &    251  &  1075\\
6   & 360   &   150  &   652 &    406  &  218 &    651  &   196  &   172 &    810 &   206  & 156  &  699 &  243 & 137 & 921  &    190  &  1011\\
7   & 2156  &   210  &  1165 &   2140  &  172 &   1002  &  1746  &   171 &   1176 &  1757  & 258  & 1350 & 1947 & 309 & 1396 &    326  &  1503\\
8   & 853   &   248  &   993 &    853  &  205 &    934  &   625  &   191 &   1110 &   632  & 221  & 1141 &  703 & 235 & 1165 &    245  &  1194\\
9   & 891   &   113  &   980 &    882  &   86 &    801  &   763  &    85 &    965 &   776  & 157  & 1195 &  907 & 226 & 1302 &    232  &  1341\\
\hline
\end{tabular}
\end{table*}

 Finally, observational incompleteness and photometric errors are
added to our synthetic CMDs, as derived from our data set, see
Sec.~\ref{sec-phot} and Sec.~\ref{sec-hst}. This allows us to compare
the observations directly with the models. As an example, the
simulated CMDs in I, I$-$K$_s$ bands of the synthetic populations are
shown in Fig.~\ref{sfh} (right panel).  Similarly to
Fig.~\ref{cmd}, only stars with V $\lsim 21$ have been plotted and
will be used in the following analysis, as they have a completeness
$>$50\% in the IR bands.

We briefly describe the main CMD features:
\begin{itemize} 
\item Model 1) from H99, it has a very well populated MS, blue loop
stars, and a compact RC where all ages are present;
\item Model 2) is very similar
to model 1,  as it is only missing its old population that does
not contribute much to the total number of stars at these magnitudes. The small
cusp in the clump region of model 1 disappears in this model; 
\item Models 3-4)  
 are missing the intermediate and young populations which are
dominant in model 1, and the models can 
be excluded by eye as a good match to the data; 
\item Models 5-6) have a very low
metallicity (Z=0.0001) and are thus bluer than model 1 and their blue
loop region is very extended; 
\item Models 7-9) with solar metallicity the RGB and the
  RC move to redder colours compared to model 1; 
\item Model 8) is very similar to model 1, and 
the only difference is that a constant SFR results in a larger
older population ($\ge$ 6 Gyr).
\end{itemize}

\subsubsection{Star counts}

\begin{figure*}
\centering
\includegraphics[width=8.5cm]{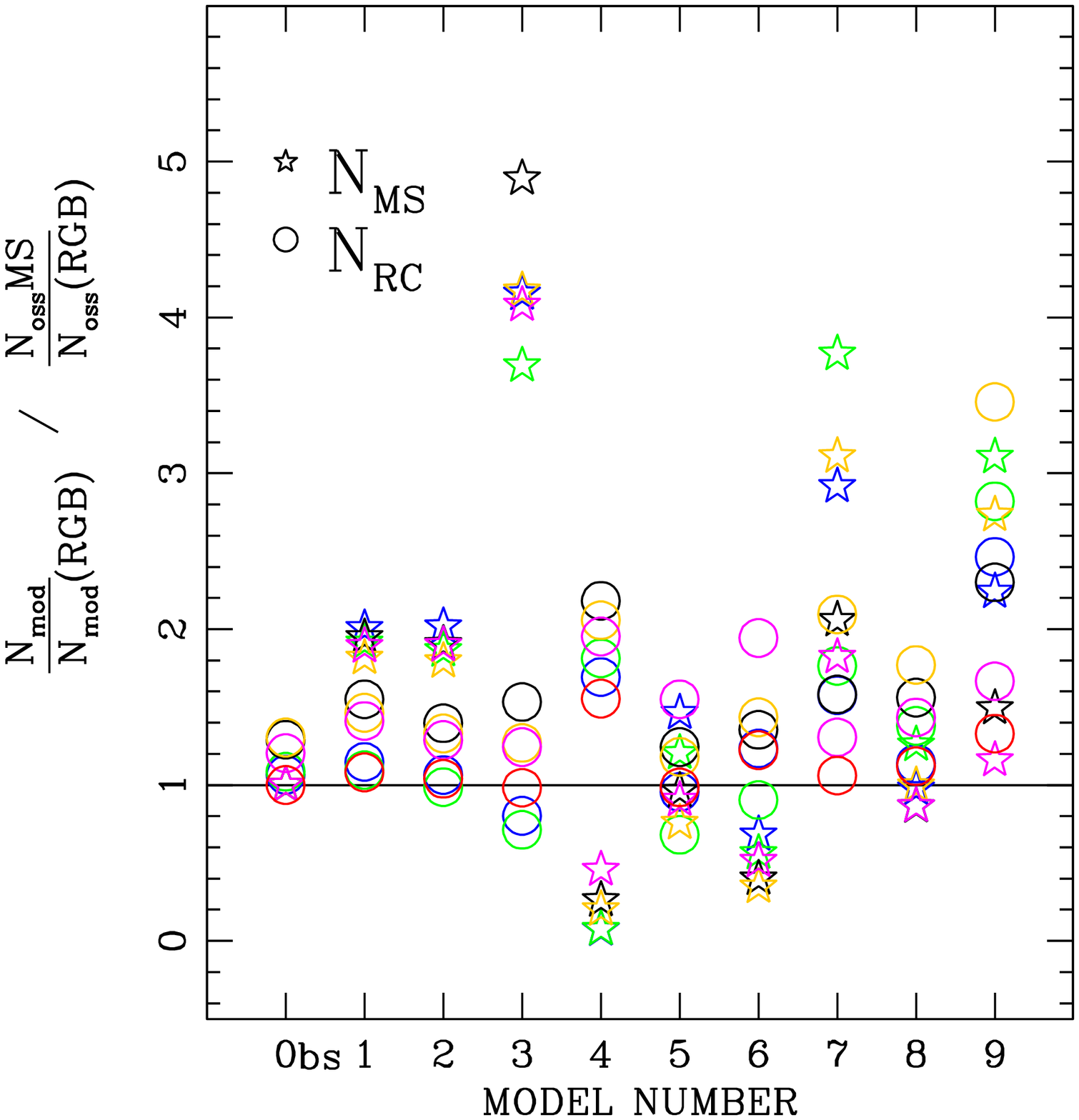}
\includegraphics[width=8.5cm]{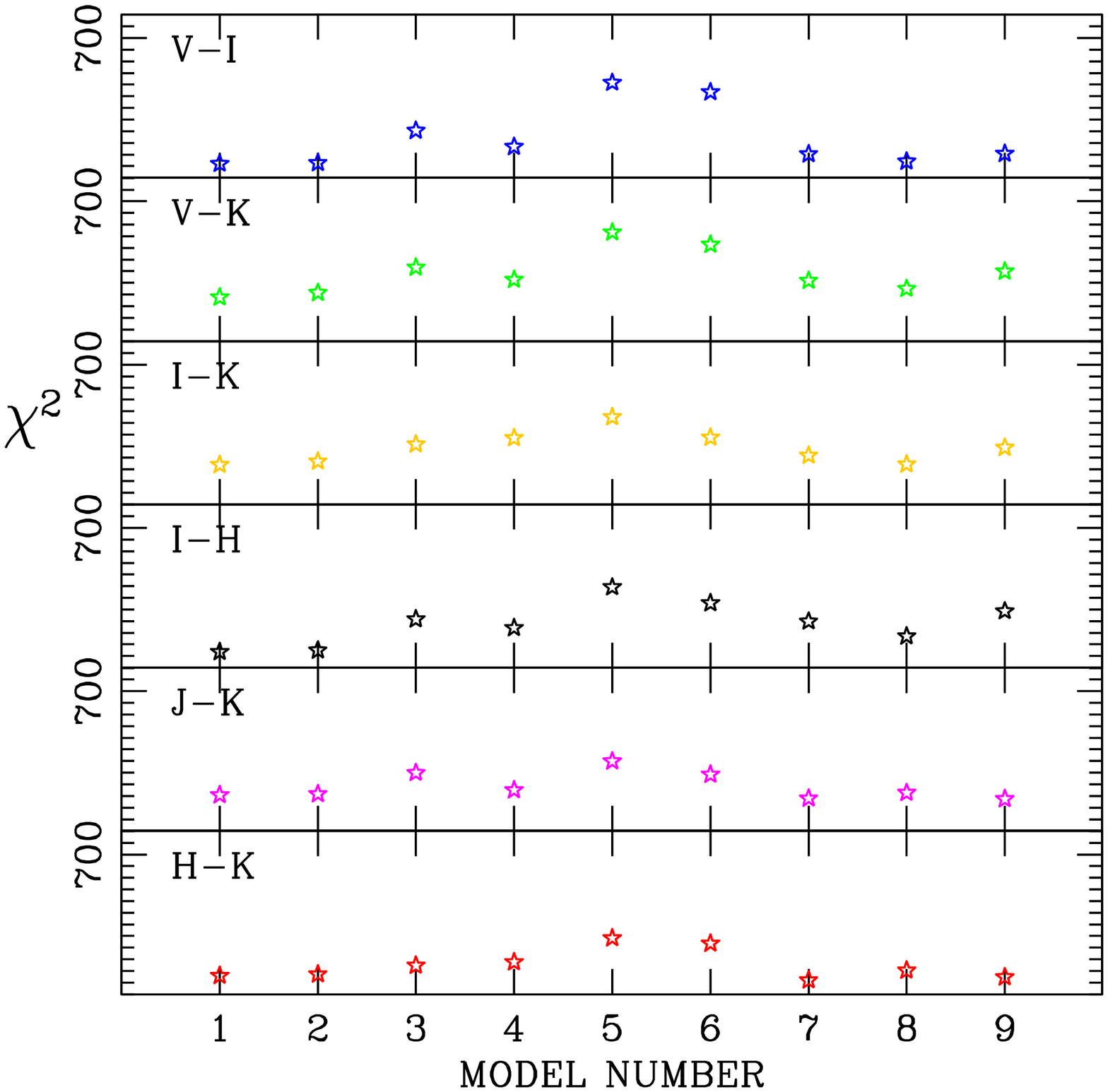}
\caption{
{\bf Left panel:} The ratio of the number of stars falling on the Main
Sequence (MS, stars) and the Red Clump (RC, circles) to those on the
RGB, as outlined in Fig.~\ref{cmd}, for both observations and models
for different filter combinations. These ratios have been
normalized to the observed MS/RGB ratio.  {\bf Right panel} The
$\chi^2$ values for different combinations of filters for the set of
models defined in Fig.~\ref{sfh}, using a bin size of 0.15 in
magnitudes and 0.25 in colour (see text for details). For this
computation we have used only the stars in F2, with V, I, J, H and
K$_s$ photometry.  {\bf Both panels:} The colour code (only in the
electronic version) is blue for V$-$I, green for V-K$_s$, yellow for
I-K$_s$, black for I-H, magenta for J-K$_s$ and red for H-K$_s$.
}\label{prob}
\end{figure*}

To compare the selected theoretical populations, shown in
Fig.~\ref{sfh} (right panel), with our observations, we first
counted the number of stars falling in the boxes (red) defined in
Fig.~\ref{cmd}. The ratio between the numbers of
RC and MS stars to RGB stars for our models (labeled 1 - 9) relative to the 
observations have been plotted in Fig.~\ref{prob} (left panel) for
different filter combinations. The observed ratios for RC and MS are roughly the
same for all the filter combinations (see also Table~\ref{table-num}), and 
an average of this (average) value is also shown (the solid line). We
expect that the best matching models are able to reproduce the same
ratio as the observations, lying very close to the solid line.
Fig.~\ref{prob} (left panel) shows that Models 8 and 5 are well
matched to observations, whereas Model 1 is not. 
This shows that with these data we are not very well able
to constrain the metallicity of the stellar population. However we are
primarily interested in the self-consistency of the predictions made
using different filter combinations.

The filter combinations shown here do give roughly the same
information, and limitations.  The only exception is the H$-$K$_s$
colour (red points in the left panel of Fig.~\ref{prob}), where the
MS is missing. The MS can produce a clear distinction between
different SFHs. However, in order to separate the MS from the RGB in
H$-$K$_s$ colour ($\le$ 0.12 mag from theoretical models) we would
need a photometric error much smaller than 0.14 mag, which is our
H$-$K$_s$ error, and it also excludes any systematics from the
calibration. The J$-$K$_s$ colour (magenta points in the
left panel of Fig.~\ref{prob}), works better, as the
colour spread is estimated to be $\sim$0.7 mag.

\subsubsection{The $\chi^2$ Test}

In this section we quantify our results using the $\chi^2$ test in the
form:
 
\begin{equation}\label{chi}
\chi ^2 = \sum{[m_\imath - n_\imath]^2/[m_\imath + n_\imath]}
\end{equation}
 
where $m_\imath$ and $n_\imath$ are the number of stars falling in the
bin $\imath$. This choice is made to account for the fact that both
the theoretical models ($m$) and the observations ($n$) are Poissonian
distributions. In fact they are both random realizations of (unknown)
distributions. As a consequence, $\sigma_{m_\imath}$=$\sqrt{m_\imath}$
and $\sigma_{n_\imath}$=$\sqrt{n_\imath}$.  We can normalize
$m_\imath$ to account for the fact that the total numbers of model and
observed stars are not the same.

The choice of the bin size is crucial and is related to the
photometric errors and the number of stars in individual bins. We
investigated the possible dependence of our results on the bin size
and adopted the most stable, which is a bin size of 0.15 in magnitude
and 0.25 in colour. Moreover, to avoid introducing spurious effects we
used only those stars with a completeness $>$50\% and observed in all
VIJHK$_s$ filters, which means field F2.

The resulting $\chi ^2$ for each colour combination is shown in
Fig.~\ref{prob} (right panel). The best models matching our data are
1, 2 and 8 for almost all the filter combinations, the exceptions are
J$-$K$_s$ and H$-$K$_s$, for which the best models are 1, 2,
7, 8 and 9. 

We can clearly see that all the colour combinations are able to
confidently exclude models 5 and 6. This result is quite
reasonable for the range of magnitudes used (V $\le$ 21), in fact we
have considered only the brightest part of the CMD which increases 
the potential degeneracies in the SFH.  Furthermore, the colour
sensitivity, estimated as the maximum amplitude in the $\chi ^2$
distribution decreases from the top to the bottom panel of Fig.~\ref{prob} (right panel), which is with increasing dependence on IR
filters.

To summarise, the V$-$I CMD shows that the maximum sensitivity to the
exact SFH comes from the HST/ACS optical data. This is mostly due to
the very small photometric errors and the high completeness of this
data set. It is also due to the greater reliability of the optical
calibration of the stellar evolutionary tracks. We must also not
forget that MAD is an experimental camera, with all the limitations
that implies.  It is nonetheless very encouraging that the optical/IR or
purely IR $\chi^2$ show the same trends, suggesting that the IR
results could be improved, with deeper more accurate photometry and
also reducing the uncertainties in the calibration of IR stellar
evolution tracks.  Thus, the standard techniques to determine the SFH
of a galaxy are also applicable to our IR data. This suggests that we
will also be able to collect useful data sets with extremely large
telescopes, using broad-band filters, from I to K$_s$.

\subsection{The Red Clump}

\begin{figure}
\includegraphics[width=8.5cm]{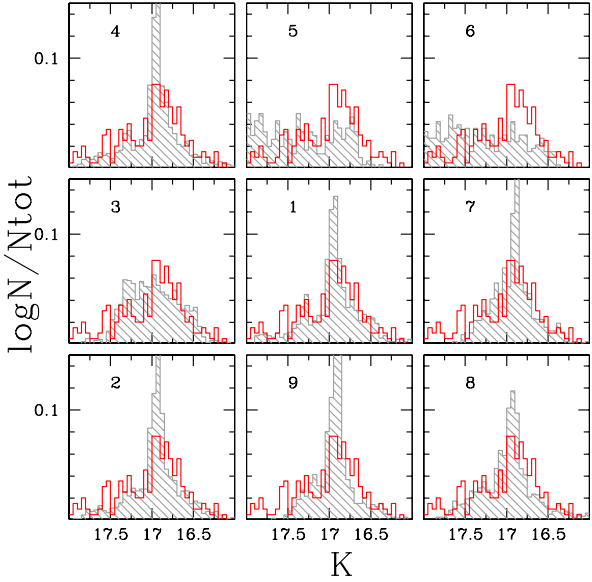}
\caption{
Observed and model luminosity functions of the red clump region in K$_s$
band, with a bin-size of 0.05 mag. The models, which are given in
Fig.~\ref{sfh} (right panel), are shown as (grey) filled histograms
and the data as (red) solid lines.
}
\label{lf_k}
\end{figure}

The RC feature is a well known and well studied distance indicator
\citep[see][and reference therein]{girardi01} in V, I and in
particular the K$_s$ filter. The distance of a galaxy based on its clump magnitude
is calibrated on nearby clump stars whose distances are known with accurate {\it Hipparcos} trigonometric
parallaxes \citep[see first row in Table~1 in][]{salaris03}. Following \citet{stanek98b}, given the histogram of
stars in the clump region per magnitude bin (see Fig.~\ref{lf_k}), the apparent clump
magnitude ($\lambda^{RC}$) is defined by performing a non linear
least squares fit of the following function:

\begin{equation}\label{eq_gauss}
N(\lambda)=a+b \lambda+c\lambda^2+d \exp{-\frac{(\lambda^{RC}-\lambda)^2}{2\sigma^2_\lambda}}
\end{equation}

The accuracy of the mean clump magnitude is limited
mainly by the photometric calibration of the data, rather than by the number
of clump stars. Furthermore, as explained in detail by 
\citet{girardi01} the main concerns in the use of RC method to
determine the distance are the extinction A$_\lambda$ and 
population effects. A measurement in K$_s$ is almost
reddening independent.

The clump region was selected in colour and magnitude as shown in
Fig.~\ref{cmd}, and the resulting K$_s$-luminosity
functions for both observation and models are shown in
Fig.~\ref{lf_k}. 
Before computing the distance with the RC method, we want to stress
that the observed luminosity function, the (red) 
solid histogram in Fig.~\ref{lf_k}, shows a
blue tail for magnitudes fainter than the main clump magnitude
($\sim$17.2 mag). This substructure in the main clump is called the ``secondary
clump''. According to \citet[][]{girardi99} this is a clear
signature of a population with age between 0.8-1.5 Gyr
(see also right panel of Fig.~\ref{sfh}). This can be particular useful when we
are not able to reach the old main sequence turn-offs, as in the
present case. To our knowledge this is the first time that a K$_s$-band
photometry is accurate enough to clearly distinguish the secondary
clump in the LMC.


The determination of the RC magnitudes are
summarised in Table~\ref{table-clump} with the corresponding
distance modulus, derived by taking into account the metallicity-population
correction given by \citet{salaris03} and assuming a reddening, E(B$-$V)$=$
0.075 mag (H99). The errors on $\lambda^{RC}$ and on $\sigma_{\lambda}$
are those given by the non-linear fit, whereas the uncertainties on the derived
distance moduli are: {\it i}) a random error which includes the errors from the
fit (0.04 mag) and the photometry  at the RC
luminosity level (0.05 mag); {\it ii}) a systematic error on the
calibration zero-point, which is non-negligible for the
K-band (0.09 mag).

\begin{table}
\caption{V, I and K$_s$ band average de-reddened magnitudes of the RC
    stars and the corresponding distance modulus obtained by accounting for the
    population effect.\label{table-clump}} 
\begin{tabular}{lccc} 
\hline
\hline
 $\lambda$ &  $\lambda^{RC}$ & $\sigma_{\lambda}$& $\mu_0$($\lambda$)\\
 &   (mag) &(mag)& (mag)\\
\hline
V &  18.98$\pm$0.04 & 0.26$\pm$0.04  &  18.51$^{a}$$\pm$0.06$_r$  \\ 
I &  18.03$\pm$0.04 & 0.28$\pm$0.03  &  18.49$^{a}$$\pm$0.06$_r$\\
K &  16.93$\pm$0.04 & 0.38$\pm$0.04  &  18.50$^{a}$$\pm$0.06$_r$$\pm$0.09$_s$\\
\hline
\end{tabular}
\\
\tablefoottext{a}{SFH by \citep{holtzman99} and the metallicity function from \citep{pagel98}, i.e. 0.26 mag, 0.20 mag and -0.03 mag for V, I and K$_s$ band respectively \citep[model Ia of Table~1 from][]{salaris03}.}\\
\end{table}

We found very similar values for the
three filters V, I and K$_s$. The average for the three filters
is $\mu_0=18.50\pm0.06_r\pm0.09_s$ mag, in excellent agreement with
previous RC-based distances
\citep{alves02,pietrzynski02,salaris03}. This is also in agreement with other independent
methods in the literature  such as Classical Cepheids, RR Lyrae and Planetary
Nebulae luminosity functions,
etc. \citep[e.g.][]{bono02,catelan08,reid10}. 
The models have been plotted in Fig.~\ref{lf_k} by
assuming a distance modulus of 18.5 mag, and they show a very good
match to our observations in particular for models 1, 2 and 8
confirming the SFH results in this section. 

\section{Discussion and Conclusions}

We have presented deep near\--IR photometry from MAD on the VLT of the
resolved stellar population in a small crowded field of the LMC bar.
Using MAD in the star oriented mode, we have reached J $\sim$ 20.6
mag, H $\sim$ 21.6 mag, and K$_s$ $\sim$ 20.6 mag with a S/N ratio $>$
10 with {\it effective} exposure times of $\sim$6~mins in J, $\sim$~36
mins in H and 20 mins in K$_s$.

The only comparable case of deep near-IR imaging of the LMC in
literature is \citet{pietrzynski02} using NTT/SOFI.  Their
observations were taken with excellent seeing conditions (DIMM
$\sim$0.6~arcsec) and at the lowest possible airmass ($\sim$1.3) from
La~Silla and over a large field of view of $\sim$5 arcmin square.
Although our optical/near-IR CMDs look deeper, thanks to the high
  resolution offered by MAD/VLT and the addition of accurate optical
  photometry, the \citet{pietrzynski02} J$-$K CMD appears to have a better
  defined RC, due most likely to the large difference in the number of stars
  observed with the bigger NTT/SOFI field of view.

Taking advantage of these MAD 
data we have also analysed the image-quality.
 It should of course be born in mind that MAD was never optimised as
a science instrument, it was always a demonstrator. 
Our main conclusions are summarised as follows:

\begin{itemize}
\item  MAD has been able to reach twice the diffraction limit in H
  and K$_s$ bands at the large zenith distances typical of the LMC. 
The maximum Strehl Ratio (SR) obtained is 
$\sim$30\%, which is better than
the expected performance for MAD in ``star-oriented''
mode. The uniformity and the stability of
  the correction varied not only with the position from the guide
  stars asterism, but also with airmass and seeing conditions. The
  complex dependency of these factors prevented us from making a direct
  comparison between our results and other MAD studies.
  However, in other experiments MAD was successfully able to reach the
  DL in K$_s$ band \citep[e.g.][]{falomo09}.

\item We quantified the constraints that appear to be necessary to push MAD
capabilities to obtain optimum results from our observations.  
It would have been 
better to observe our target with an airmass $<$1.5 and DIMM less
than $<$1~arcsec {\it to obtain results within at least a factor two of the
diffraction limit for this system}. However it should also be noted
that we see a significant improvement to natural seeing (around a
  factor 2) even for the
unfavourable conditions under which the LMC was sometimes observed. 

\item Another fundamental requirement for deep AO imaging 
is the ability to effectively
co-add large numbers of measurements (made on individual images) to
build up a final product that reaches faint magnitude limits with
an angular resolution comparable to the diffraction limit. With
current standard data reduction, image analysis and correction
techniques we had to throw away a large fraction of our images
(e.g. $\sim 66\%$ in K$_s$ filter). This could presumably 
be improved upon with 
clever post-processing techniques to correct
images before analysing them. This may be achieved with the aid of PSF
reconstruction techniques which should provide an accurate theoretical
understanding of the form and variation of the PSF depending upon the
atmospheric and instrumental variations. 
This would hopefully mean
that less data need to be discarded during reduction and analysis, and
thus providing deeper and sharper images.
\item The major and apparently unique advantage of MCAO, is 
that it can obviously provide a uniform AO correction over a wide field, 
allowing the straight forward use of 
the standard PSF modeling and fitting procedures.

\item In our simplified scenario we assume that our data are
identical to previously LMC bar results 
  \citep{holtzman99}. Then, using our final optical/IR catalogue (VIJHK$_s$), where the optical
observations come from ACS/HST, we could determine how well stellar
evolution models in the different filter combinations match our data
for an assumed SFH \citep{holtzman99}. 
We conclude that we can be confident that the optical 
techniques for determining SFHs 
in V and I, can also be applied to IR data
sets.
Of course improvements in the calibration of IR isochrones would also be
very helpful and are still needed.
\item From our RC analysis we derive a new, accurate, distance to the
  LMC: $\mu_0=18.50\pm0.06_r\pm0.09_s$ mag, assuming E(B$-$V)$=$ 0.075 mag. An inspection of the RC luminosity
  function confirms the evidence of the ``secondary clump'' feature in the K$_s$-band. 

\end{itemize}

In conclusion, MAD is a demonstrator instrument with a small engineering grade detector, and yet the experiment worked very well. The future E-ELT will sample the atmospheric 
turbulence with a sampling step two times finer than MAD on VLT. This is just to stress that the peak
correction achievable with MAD is smaller than that expected
from the E-ELT. Thus some of the technical
problems that we had 
with MAD, we would not expect with the E-ELT that will be much more
stable in terms of uniformity and performance \citep[e.g.,][]{deep11}.

\section*{Acknowledgments}
 
We would like to thank the MAD team for useful suggestions during
the preparation and carrying out of our observations, with
particularly thanks to P. Amico and 
E. Marchetti. GF thanks M. Monelli for interesting
discussions and for his help with IAC-STAR code, which has been used
in this work. IAC-STAR is supported and maintained by the computer
division of the Instituto de Astrofisica de Canarias.
We thank the anonymous referee for invaluable help 
clarifying the results presented in this paper.

This publication makes use of data products from the Two Micron All
Sky Survey, which is a joint 
project of the University of Massachusetts and Infrared Processing
and Analysis Center/California Institute of Technology,
funded by the National Aeronautics and Space Administration
and the National Science Foundation. 
ET \& GF have been supported by an NWO-VICI grant.

\bibliographystyle{aa} 
\bibliography{mad}

\end{document}